\documentclass[prl,twocolumn,showpacs,amsmath,amssymb]{revtex4}

\usepackage{graphicx}
\usepackage{dcolumn}
\usepackage{bm}

\newcommand{\siml}{\raisebox{-.6ex}{$\stackrel{<}{\displaystyle{\sim}}$}}

\begin{document}

\title{A Solvable Sequence Evolution Model and Genomic Correlations}

\author{Philipp W. Messer$^1$, Peter F. Arndt$^2$, and Michael L\"assig$^1$}
\affiliation{$^{1}$Institute for Theoretical Physics, University of Cologne, Z\"ulpicher Str.~77, 50937 K\"oln, Germany}
\affiliation{$^{2}$Max Planck Institute for Molecular Genetics, Ihnestr.~73, 14195 Berlin, Germany}

\date{\today}

\begin{abstract}
  We study a minimal model for genome evolution whose elementary
  processes are single site mutation, duplication and deletion of
  sequence regions and insertion of random segments. These processes
  are found to generate long-range correlations in the composition of
  letters as long as the sequence length is growing, i.e., the
  combined rates of duplications and insertions are higher than the
  deletion rate. For constant sequence length, on the other hand, all
  initial correlations decay exponentially. These results are obtained
  analytically and by simulations. They are compared
  with the long-range correlations observed in genomic DNA, and the
  implications for genome evolution are discussed.
\end{abstract}

\pacs{87.23.Kg, 87.15.Cc, 05.40.-a}

\maketitle

Over a decade ago, long-range correlations in the sequence composition
of DNA have been discovered~\cite{Peng92,Voss92,Li92}. With the
rapidly growing availability of whole-genome sequence data, the
composition of genomic DNA can now be studied systematically over a
wide range of scales and organisms.  The statistical analysis is quite
intricate since genomic DNA is a rather ``patchy'' statistical
environment~\cite{Karlin93}: it consists of genes, noncoding regions,
repetitive elements etc., and all of these substructures have a
systematic influence on the local sequence composition. Variations in
composition along the genome have been studied extensively by a number
of different
methods~\cite{Li97,Vieira99,Bernaola02,Arneodo95,Peng94,Stanley99,Holste03,Ouyang04}, and it is now well established that
long-range correlations in base composition appear in the genomes of
many species.  These can be measured, for example, by the
autocorrelation function $C(r)$ of the GC-content, which measures the
likelihood of finding G-C Watson-Crick pairs at a distance of $r$
bases along the backbone of the DNA molecule.  However, the form of
these correlations is much more complex than simple power laws. Within
one chromosome, there is often a variety of different scaling regimes
and effective exponents, and sometimes no clear scaling at all.
Moreover, the effective exponents of comparable scaling regimes vary
considerably between different species, and even between different
chromosomes of the same species~\cite{Bernaola02,Holste03,next}.

Despite the ubiquity of genomic correlations, little is known about
their evolutionary origin. In this Letter, we address the question
whether the observed correlations can be explained quantitatively by a
biologically realistic ``minimal'' model of sequence evolution. We
take into account four well known elementary evolutionary modes:
single site mutations, duplications and deletions of existing segments of the 
sequence, and insertions of random segments. The duplication
processes are believed to be a crucial mechanism of genome
growth~\cite{Goffeau04,Eichler02a,Hsieh03}; the length of the
duplicated segments ranges from single letters to thousands of letters
as in the case of gene duplications. The model is minimal in the sense
that all four elementary modes are {\em local} stochastic processes
compatible with {\em neutral evolution}, i.e., they do not require any
assumption of natural selection.  An alternative possible reason for
the observed correlations may be {\em long-range interactions} likely
to be caused by natural selection for a specific local GC-content. An
example of such a selective process is the clustering of genes in some
regions of a chromosome~\cite{Lercher03}, but no plausible
mechanism producing  long-range interactions has been
proposed so far.

Li's original work has shown that already a simple stochastic process
consisting of duplications and mutations of single letters leads to
generic power law correlations in the sequence
composition~\cite{Li91}. Here we analyze in detail the generalized
sequence evolution model introduced above. In particular, we calculate
the stationary two-point correlation function $C(r)$.  It is of power
law form, $C(r) \sim r^{-\alpha}$, with a decay exponent $\alpha$
depending on only two effective parameters, which are simple functions
of the rates of the elementary processes.  These long-range
correlations are generic as long as the rates of the processes result
in a growing sequence. At constant sequence length, however, the
stationary correlations in sequence composition vanish, and initial
correlations from a previous growth phase decay. Our analytic results
(which differ from Li's approximate expressions~\cite{Li91} and the
results of~\cite{Mansilla00}) are in excellent agreement with our
numerical simulations.  We use these results to infer from measured
values of $\alpha$ a lower bound on the growth rate of the genome,
which can be compared with independent estimates. The implications of
our findings on the evolution of mammalian genomes are discussed at
the end of this Letter.  
%\medskip

{\em Sequence evolution model.}---The stochastic evolution model
generates sequences $(s_1, \dots, s_N)$ of variable length $N$. For
simplicity, their letters are taken from a binary alphabet; $s_k =
\pm1$. (In the application to genomic systems, $s_k = +1$ denotes a
GC-pair and $s_k = -1$ an AT-pair at backbone position $k$.) The
elementary evolutionary steps are mutations, duplications, insertions,
and deletions of single letters (the generalization to segments will
be discussed below). They are Markov processes with rates $\mu$,
$\delta$, $\gamma^+$, and $\gamma^-$ acting on the sequences as
\begin{equation}
\label{rep_mut_processes}
(\cdots,s,s',\cdots)\to\left\{\begin{array}{l@{\;\;:\;\;}l} 
(\cdots,-s,s',\cdots)&\mbox{rate }\mu \\
(\cdots,s,s,s',\cdots)&\mbox{rate }\delta \\
(\cdots,s,x,s',\cdots)&\mbox{rate }\gamma^+\\
(\cdots,s',\cdots)&\mbox{rate }\gamma^-,
\end{array}\right.
\end{equation}
where $x = \pm 1$ denotes an uniformly distributed random letter.
Duplication and insertion events introduce a new letter next to an
exiting one and shift all subsequent letters one position to the
right, thereby increasing the sequence length by~1. Conversely,
deletions shorten the length by~1. This type of Markov evolution model
is widely used in computational biology, forming the statistical basis
of sequence alignment algorithms~\cite{Durbin98}. Running all four
processes over a time $t$ produces a statistical ensemble of
sequences; the corresponding averages are denoted by $\langle \dots
\rangle (t)$. This ensemble is characterized by the rates $\delta$,
$\mu$, $\gamma^+$, $\gamma^-$, and by the initial sequence. Here we
use sequences of length $1$ with a fixed letter, $(s_1)=1$, or a
random letter, $(s_1) = x$.

After a time $t$, the sequences have an average length $\langle N
\rangle(t) = \exp(\lambda t)$ with the effective growth rate
\begin{equation}
\lambda = \delta+\gamma^+-\gamma^-.
\label{lambda}
\end{equation}
We are interested in two dynamical regimes, sequence growth from a
single-letter initial state (i.e., $\lambda > 0$) and the evolution of
sequences at stationary length $\langle N\rangle \gg 1$ (i.e.,
$\lambda = 0$), to which we now turn in order.  
%\medskip

{\em Growth dynamics and stationary correlations.}---The composition
bias of the sequences at position $k$ is measured by the expectation
value $\langle s_k\rangle(t)$. It is easy to show that any initial
composition bias decays due to mutations and random insertions. We
note that each insertion can be regarded as a duplication with a
subsequent mutation in half of the cases, resulting in an effective
mutation rate
\begin{equation}
\mu_{\rm eff}=\mu+\gamma^+/2.
\label{mu_eff}
\end{equation}
We obtain $\langle s_k\rangle (t)\propto\exp(-2\mu_{\rm eff}t)$ for
fixed initial condition, while $\langle s_k\rangle (t)=0$ for random
initial conditions. 
The composition correlation $C(r) \equiv \langle s_k s_{k+r} \rangle
(t)$ between two sequence positions at distance $r$ is affected by all
four processes and is independent of the initial condition. Its
evolution equation can be derived by writing it as $C(r,t)=P_{\rm
  eq}(r,t)-P_{\rm op}(r,t)$, where $P_{\rm eq}(r,t)$ and $P_{\rm
  op}(r,t)$ denote the joint probabilities of finding two symbols of
equal and opposite signs, respectively, at a distance $r$. The Master
equation for $P_{\rm eq}(r,t)$ takes the form
\begin{eqnarray} 
\frac{\partial}{\partial t} P_{\rm eq}(r,t)&=&
2\mu_{\rm eff}\:[-P_{\rm eq}(r,t)+P_{\rm op}(r,t)]
\nonumber\\ 
&&+[r\delta+(r-1)\gamma^+]\:[P_{\rm eq}(r-1,t)-P_{\rm eq}(r,t)]
\nonumber\\ 
&&+r\gamma^-\:[P_{\rm eq}(r+1,t)-P_{\rm eq}(r,t)]. 
\end{eqnarray} 
The first term on the r.h.s.~describes the change in $P_{\rm eq}(r,t)$
due to mutations and random insertions, while the second term
specifies the probability current due to duplication of a site in the
interval $(k,k+r-1)$ or insertion of a new site in the interval
$(k,k+r-2)$. The third term gives the corresponding current due to
deletions. By exchanging $P_{\rm eq}$ and $P_{\rm op}$, we obtain a
similar equation for $P_{\rm op}(r,t)$. Hence we have
\begin{eqnarray}
\label{master_equation_C}
\frac{\partial}{\partial t} C(r,t)&=&-4\mu_{\rm eff}\:C(r)\nonumber\\
&&+[r\delta+(r-1)\gamma^+]\:[C(r-1)-C(r)]\nonumber\\
&&+r\gamma^-\:[C(r+1)-C(r)]. 
\end{eqnarray} 
For the special case with only single-letter duplications and
mutations ($\delta,\mu > 0$, $\gamma^+=\gamma^-=0$), which is
equivalent to Li's original model~\cite{Li91}, we find a simple
analytical form for the stationary $C(r)$ by solving the recursion
\begin{equation} 
\label{recursion_C}
C(r)=\frac{r}{\alpha+r}\:C(r-1)\quad\mbox{with}
\quad\alpha=\frac{4\mu}{\delta}
\end{equation} 
and the initial value $C(0)=1$. This gives
\begin{equation} 
\label{C_l_result}
C(r)=\frac{\Gamma(r+1)\Gamma(1+\alpha)}{\Gamma(r+1+\alpha)}=\frac{\alpha}{1+\alpha}\:B(r,\alpha),
\end{equation} 
where $\Gamma(x)$ is the gamma function and $B(x,y)$ the beta
function.  Evaluating its asymptotic behavior for $x\gg1$,
\begin{equation}
B(x,y)\propto\Gamma(y)\:x^{-y}
\left[1-\frac{y(y-1)}{2x}\left(1+\mbox{O}
\left(\frac{1}{x}\right)\right)\right]\nonumber,
\end{equation} 
then produces the algebraic decay $C(r)\propto r^{-\alpha}$. For the
general case including insertions and deletions, the asymptotic decay
can still be obtained exactly in the continuum limit. For $r \gg 1$
and $\delta > 0$, the difference equation~(\ref{master_equation_C})
becomes the differential equation
\begin{equation}
\label{C_l_indel_dgl}
\frac{\partial}{\partial t} C(r,t)=-4\mu_{\rm eff}C(r,t)-r\lambda
\frac{\partial}{\partial r} C(r,t)
\end{equation}
with the effective rates $\mu_{\rm eff}$ and $\lambda$ defined by 
(\ref{lambda}) and (\ref{mu_eff}). This
has the stationary solution 
\begin{equation}
\label{C_l_indel_asymptotics}
C(r)\propto r^{-\alpha}\quad\mbox{with}
\quad\alpha=\frac{4\mu_{\rm eff}}{\lambda}. 
\end{equation} 
%\medskip

Eq.~(\ref{C_l_indel_dgl}) clearly shows the mechanism generating
long-range correlations in this type of sequence evolution model.
Correlations are continuously produced at small scales by duplications
and transported to larger distances by the net exponential expansion
of the sequence (resulting from duplications and
insertions/deletions). On the other hand, correlations decay
exponentially due to processes randomizing the sequence (i.e.,
mutations and random insertions). The competition between expansion
and randomization produces the algebraic decay $C(r)\propto
r^{-\alpha}$, which is highly universal.  Microscopic details of the
evolution processes are irrelevant, the exponent $\alpha$ is
determined by a simple balance between the growth rate $\lambda$ and
the effective mutation rate $\mu_{\rm eff}$. Hence, an extended model
containing duplications, deletions and random insertions of sequence
{\em segments} of finite length $\ell= 1,2,...,\ell_{\rm max}$ with
respective rates $\delta_l$, $\gamma_{\ell}^-$, and $\gamma_{\ell}^+$
still has the same asymptotics ~(\ref{C_l_indel_asymptotics}) for
$N(t)\gg\ell_{\rm max}$ and $r\gg\ell_{\rm max}$. The effective rates
(\ref{lambda}), (\ref{mu_eff}) are now given by
\begin{equation}
\label{full_effective_rates}
\lambda=\sum_{\ell}\ell\,\left[\delta_{\ell}+\gamma^+_{\ell}-\gamma^-_{\ell}\right],\quad\mu_{\rm eff}=\mu+\frac{1}{2}\sum_{\ell}\ell\gamma_{\ell}^+.
\end{equation} 
This asymptotics can again be proved from an exact Master equation
similar to (\ref{master_equation_C})~\cite{next}.  The extended model
is important for genomic evolution since strong long-range
correlations (i.e., small values of $\alpha$) can be the combined
result of segment duplications with different values of $\ell$. Their
individual rates $\delta_\ell$ might be small and difficult to assess
but the cumulative rate $\lambda$ can still be estimated.
%\medskip

\emph{Stationary-length dynamics and time-dependent
  correlations.}---It is obvious from Eq.~(\ref{C_l_indel_dgl}) that
stationary long-range correlations only exist as long as the sequence
grows, i.e. for $\lambda > 0$. Consider now the following evolutionary
scenario: sequence growth with rate $\lambda_1 > 0$ up to a length
$N_0 = N(t_0)$, followed by a second phase with $\lambda_2 = 0$ and
$\langle N \rangle(t) = N_0$ for $t > t_0$. The time-dependent
solution of Eq.~(\ref{C_l_indel_dgl}) for the asymptotics of $C(r,t)$ is
then 
\begin{equation}
\label{C_l_theta_deletion}
C(r,t) = C(r,t_0)\:e^{-4\mu_{\rm eff}\Delta t} 
\propto r^{-4 \mu_{\rm eff}/\lambda_1} e^{-4\mu_{\rm eff}\Delta
t}
\end{equation}
with $\Delta t=t-t_0>0$. In the second phase, the long-range tails of
$C(r,t)$ are preserved but their amplitude decays with a
characteristic time scale $\tau=(4\mu_{\rm eff})^{-1}$.
%\medskip

\emph{Numerical results.}---We have performed extensive Monte Carlo
simulations of our model. During each time step $\Delta
t=[(\mu+\sum\nolimits_{\ell}[\delta_{\ell}+\gamma_{\ell}^++\gamma_{\ell}^-])N(t)]^{-1}$
we choose a random site and apply one of the elementary processes with
its relative weight.
\begin{figure} [t!]
\centering
\includegraphics[width=0.92\linewidth]{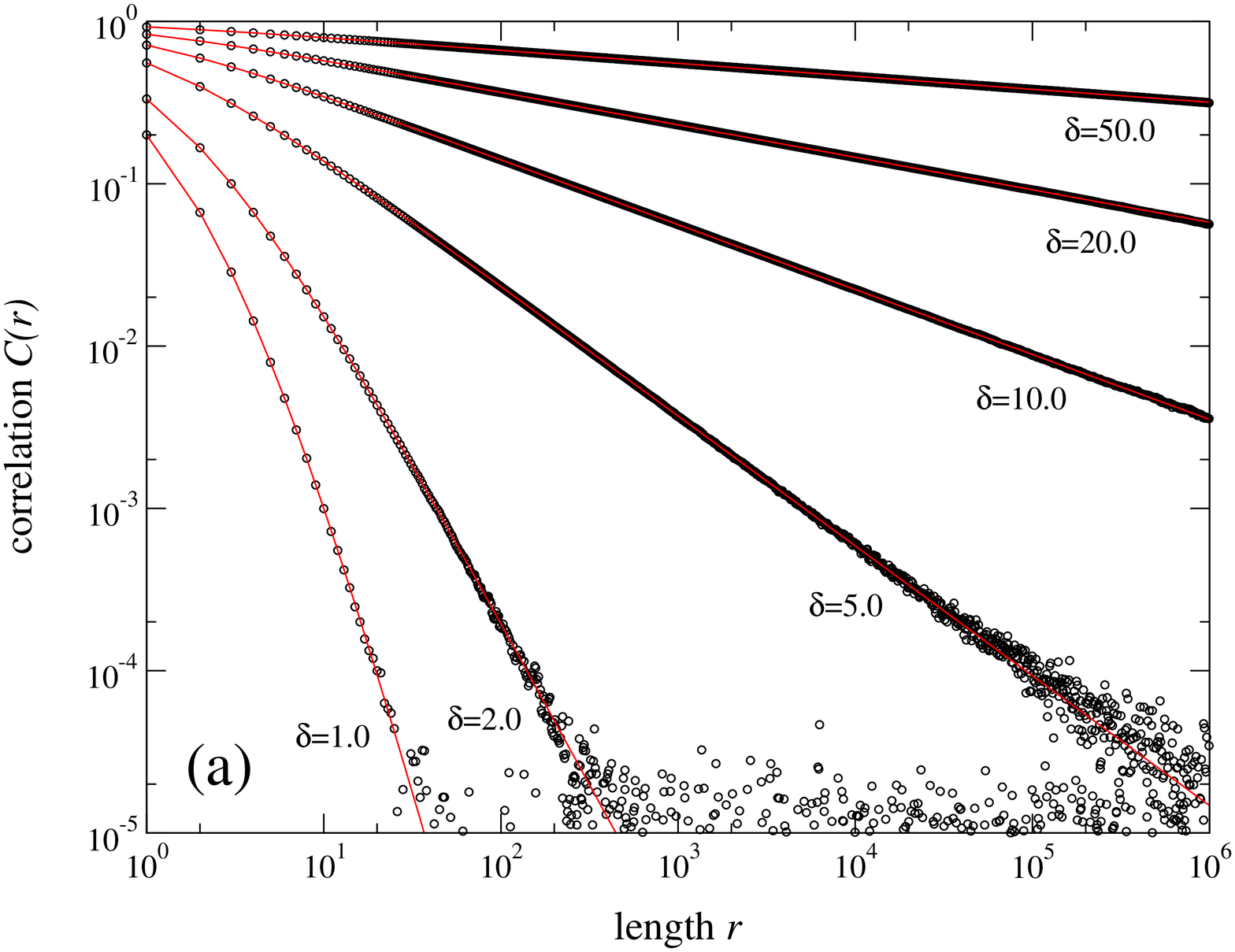}
\includegraphics[width=0.92\linewidth]{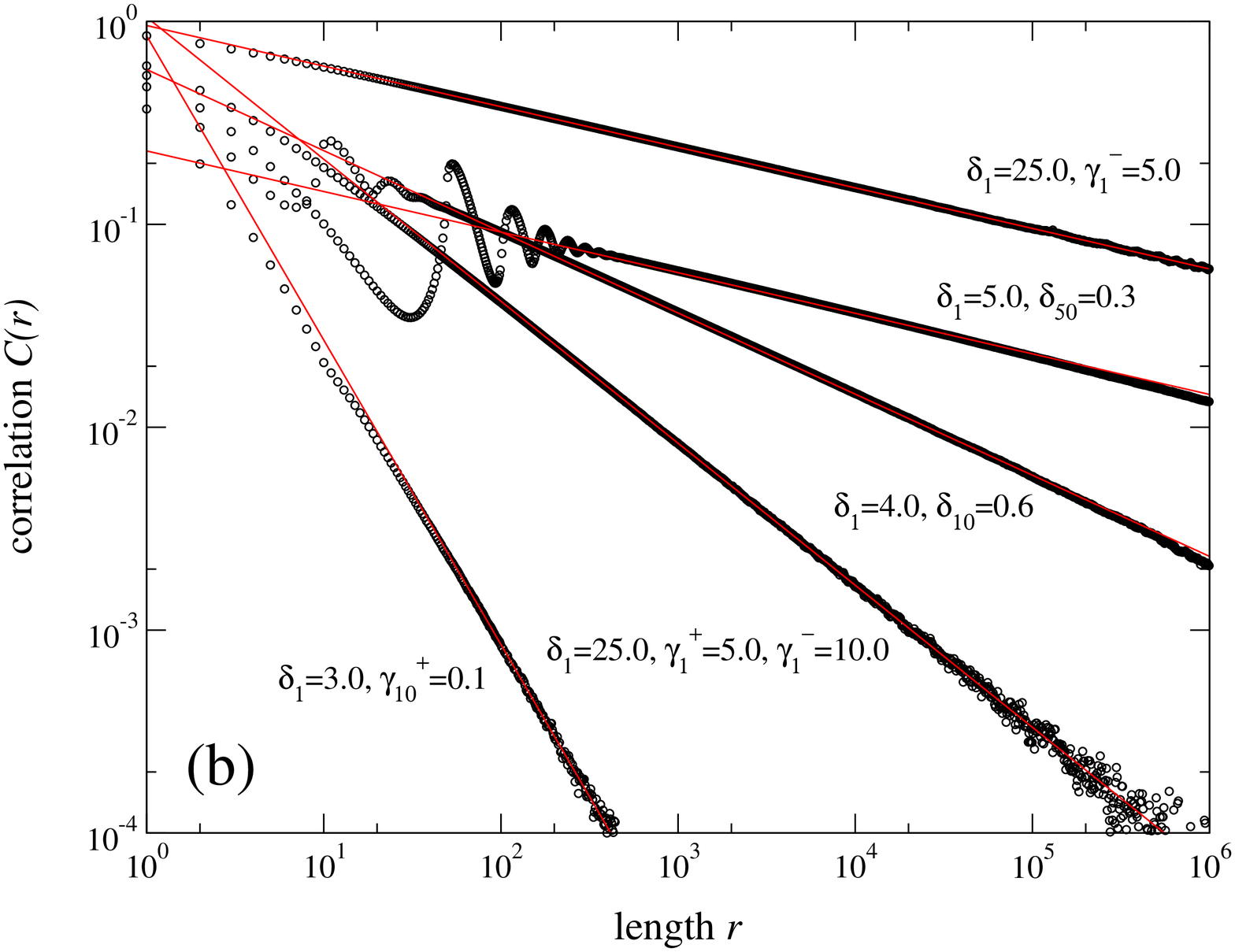}
\caption{Stationary $C(r)$ at different rates of the
  elementary processes. (a) Single-letter duplication-mutation model:
  Numerical results (circles) and the analytical
  form~(\ref{C_l_result}) (lines) for $\mu=1$, $\delta$ varying. (b)
  Full model: Numerical results (circles) with the analytic
  asymptotics~(\ref{C_l_indel_asymptotics})
  and~(\ref{full_effective_rates}) (lines) for $\mu=1$ and varying
  rates of the other processes (rates not specified in the plot are
  zero). The dynamics of the sequences was simulated until they
  reached a length of $N=2^{27}\approx10^8$; $C(r)$ was averaged over
  the sequence and over 100 runs.
\label{cor_eps}}
\end{figure}
For a single realization of this dynamics, the correlation function
$C(r)$ is well approximated by the sequence average $(N-r)^{-1}
\sum_{k=1}^{N-r} s_k s_{k+r}$. Further averaging over 100 realizations
produces very accurate measurements of $C(r)$.

Fig.~\ref{cor_eps}(a) shows the numerical $C(r)$ for the single-letter
duplication-mutation dynamics with various rates, which is in
excellent agreement with the analytic expression~(\ref{C_l_result}).
The same is shown in Fig.~\ref{cor_eps}(b) for the general case with
all types of processes present, verifying the asymptotic
behavior~(\ref{C_l_indel_asymptotics})
and~(\ref{full_effective_rates}). For completeness, we have also
obtained power spectra and the mutual information function, as defined
in~\cite{Holste03}, which have the expected decay exponents $1
-\alpha$ and $2\alpha$, respectively.

\begin{figure} [t!]
\centering
\includegraphics[width=0.92\linewidth]{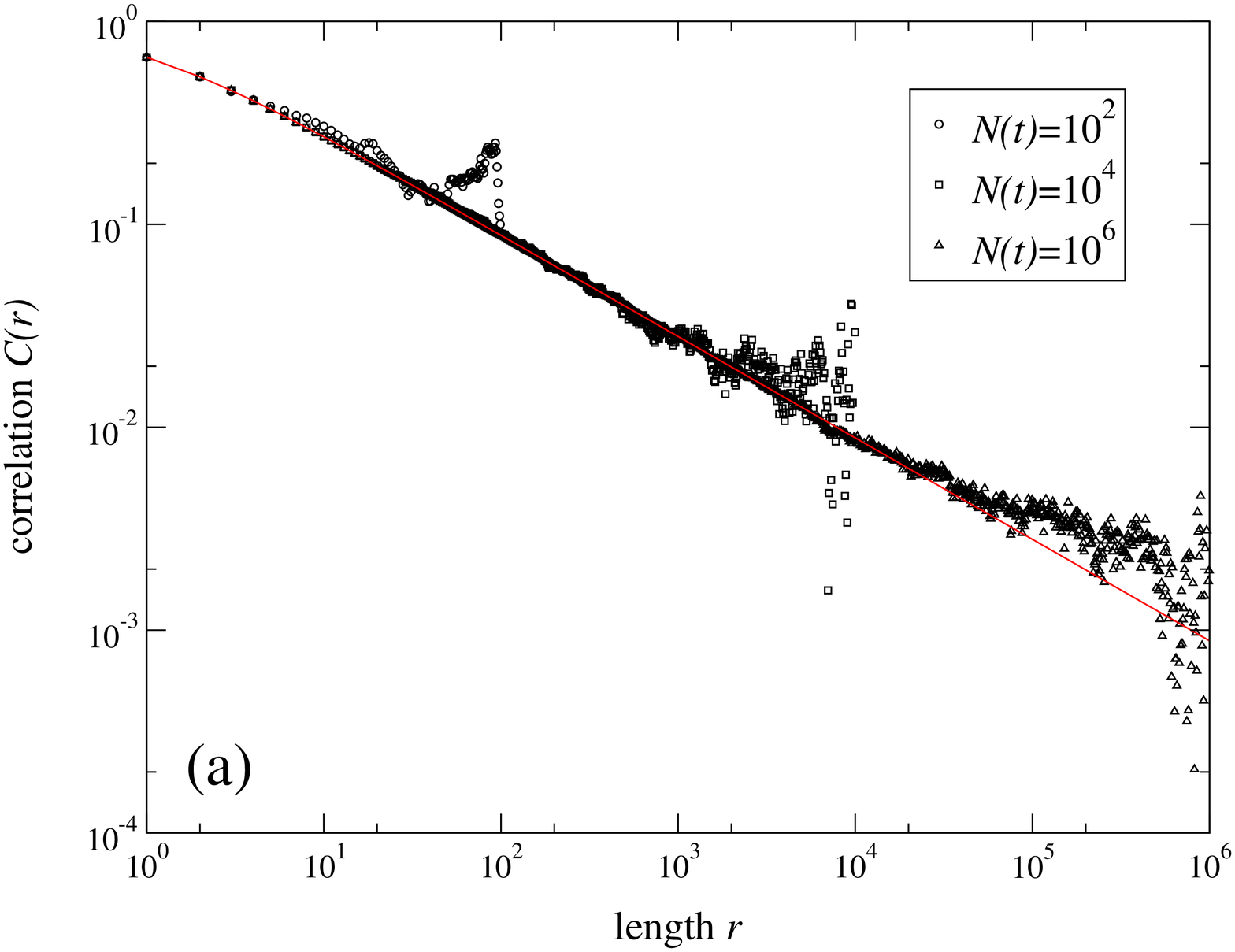}
\includegraphics[width=0.92\linewidth]{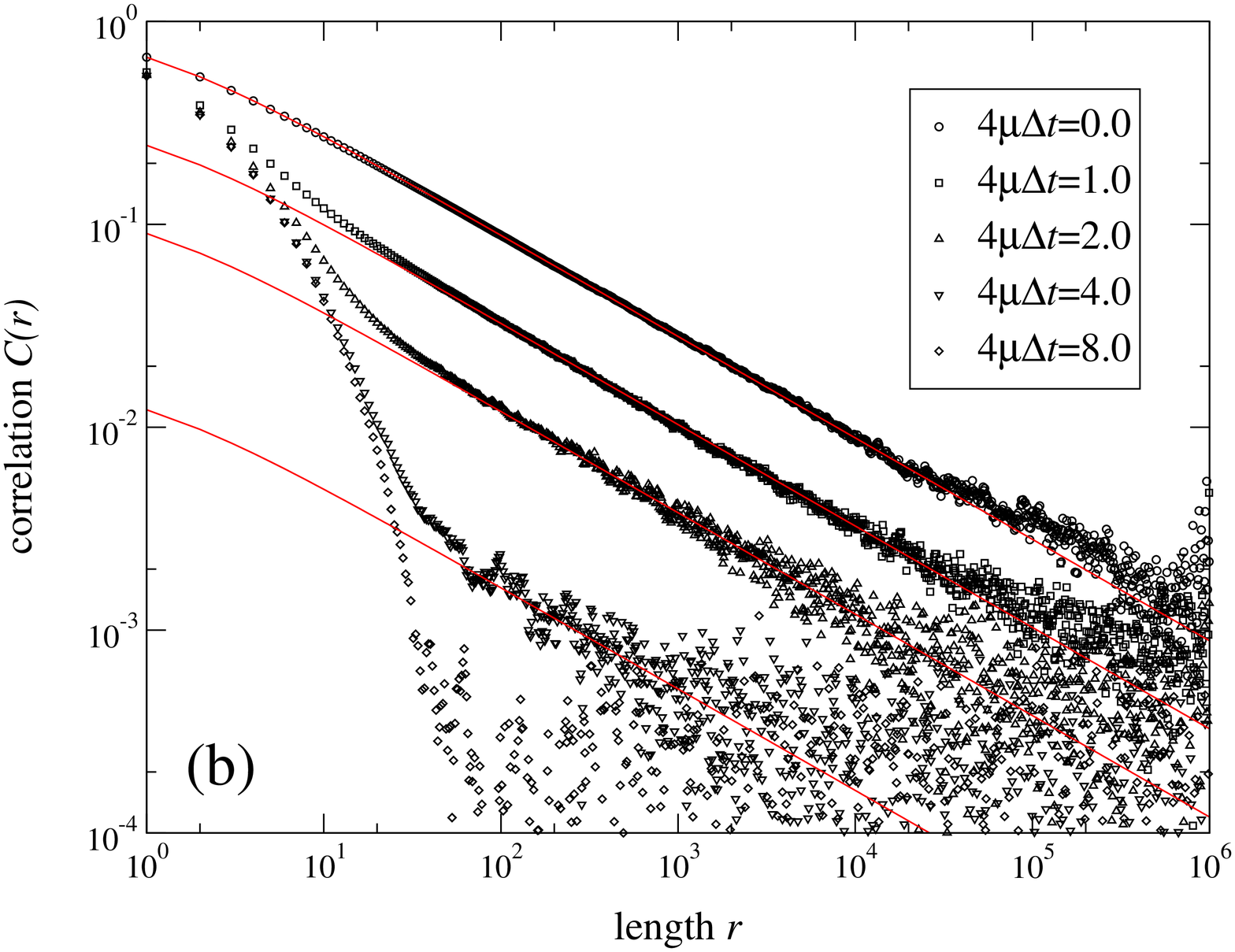}
\caption{Time-dependent correlations $C(r,t)$. (a) Build-up of long-range
  correlations by stationary growth. Measured $C(r,t)$ at various
  intermediate lengths $N(t)=10^2,10^4,10^6$ (symbols) together with
  the stationary form~(\ref{C_l_result}) (line) for $\mu=1$,
  $\delta_1=\delta=8$, all other parameters are zero. (b) Decay of
  correlations during sequence evolution at stationary length
  $N_0=10^6$. Measured $C(r,t)$ at various times $\Delta t$ (symbols)
  together with the analytic decay of the long-range tail given by
  Eq.~(\ref{C_l_theta_deletion}) (lines). Note that there are still
  correlations remaining on short length scales.
\label{growth_decay_eps}}
\end{figure}

The dynamical build-up of these correlations for growing sequences is
seen in Fig.~\ref{growth_decay_eps}(a), which shows $C(r,t)$ at
various intermediate times of the growth process. The correlation
rapidly converges to the stationary form for all distances
$r\:\siml\:N(t)$.  This should be compared with the time-dependence of
$C(r,t)$ at constant length in Fig.~\ref{growth_decay_eps}(b), which
shows an algebraic tail with an exponentially decreasing amplitude as
predicted by Eq.~(\ref{C_l_theta_deletion}).
%\medskip

{\em Genomic evolution.}---As pointed out above, the processes
discussed here build a minimal model for dynamically generated
long-range correlations along a sequence. But can this model explain
the observed correlations in genomic DNA?  The correlation function
$C(r)$ along human chromosomes shows a rather slow algebraic decay on
distance scales $10^3<r<10^6$ with typical effective exponents
$\alpha\approx0.1$~\cite{Bernaola02,Holste03}. We have confirmed these
measurements and found them to be consistent with sequence data from
other mammals~\cite{next}. A lower bound of the effective mutation
rate in mammals is $\mu_{\rm eff}\approx 2\cdot10^{-9}\rm a^{-1}$ per
site~\cite{Arndt04}.  Assuming stationary growth, we can use these
values of $\alpha$ and $\mu_{\rm eff}$ to derive a lower bound on the
genomic growth rate $\lambda$, resulting in a minimum value $\lambda
\approx 10^{-7}\rm a^{-1}$ per site according to
Eq.~(\ref{C_l_indel_asymptotics}).  However, this rate is much too
high. Our genome would have expanded much faster than it is observed
since the current human genome contains $N\approx3\cdot 10^9$ base
pairs and, assuming the above rate of genome expansion, would have
contained only about $4\cdot 10^5$ base pairs at the time of mammalian
radiation about 90 million years ago. This can clearly be rejected
since approximately 40\% of the human genome can be aligned to the
mouse genome, representing most of the orthologous sequences that
remain in both lineages from the common ancestor~\cite{Mouse02}.

Over longer evolutionary periods, genomic expansion phases with rates
$\lambda \sim 10^{-7}\rm a^{-1}$ cannot be ruled out if we assume the
history of the genome has been a {\em punctuated} process, with such
expansion phases followed by periods of approximately constant length.
In the human genome, there is by now ample evidence for growth by
segmental duplications with various segment
lengths~\cite{Thomas04,Eichler02b}. In a punctuated growth process,
correlations are produced and transported during the expansion phases.
During the stationary phases, the previously established correlations
decay as given by Eq.~(\ref{C_l_theta_deletion}). In mammals, the last
period of rapid expansion has been the mammalian radiation, and the
characteristic time scale of the decay is $\tau\approx100$~Myr.
Correlations present or generated at the time of the mammalian
radiation would hence still persist. The succession of several
distinct growth phases with different values of $\lambda$ and
$\mu_{\rm eff}$ could even explain correlations $C(r)$ with several
scaling regimes as found in human chromosomes~\cite{Bernaola02}. We
conclude that the correlations observed in mammals are compatible with
a punctuated expansion-randomization process. Of course, this does not
rule out other causes. Indeed, the rather diverse functional forms
found in different species may point towards more than one generating
mechanism. If genomic expansion proves to be a significant
contribution, composition correlations could be the ``background
radiation'' of genomics, allowing us to trace the history of genomes
far back in evolutionary time.


\begin{references}

\bibitem{Li92}
W. Li and K. Kaneko,
{\it Europhys. Lett. }{\bf 17}, 655 (1992).

\bibitem{Peng92}
C.-K. Peng {\it et al.},
{\it Nature }(London) {\bf 356}, 168 (1992).

\bibitem{Voss92}
R. F. Voss,
{\it Phys. Rev. Lett. }{\bf 68}, 3805 (1992).

\bibitem{Karlin93}
S. Karlin and V. Brendel,
{\it Science }{\bf 259}, 677 (1993).

\bibitem{Peng94}
C.-K Peng {\it et al.},
{\it Phys. Rev. E }{\bf 49}, 1685 (1994).

\bibitem{Arneodo95}
A. Arneodo, E. Bacry, P. V. Graves, and J. F. Muzy,
{\it Phys. Rev. Lett. }{\bf 74}, 3293 (1995). 

\bibitem{Li97}
W. Li,
{\it Comput. Chem. }{\bf 21}, 257 (1997).

\bibitem{Vieira99}
M. de Sousa Vieira,
{\it Phys. Rev. E }{\bf 60}, 5932 (1999).

\bibitem{Stanley99}
H. E. Stanley {\it et al.},
{\it Physica A }{\bf 273}, 1 (1999).

\bibitem{Bernaola02}
P. Bernaola-Galvan, P. Carpena, R. Roman-Roldan, and J. L. Oliver,
{\it Gene }{\bf 300}, 105 (2002). 

\bibitem{Holste03}
D. Holste {\it et al.},
{\it Phys. Rev. E }{\bf 67}, 061913 (2003).

%\bibitem{Isohata03}
%Y. Isohata and M. Hayashi,
%{\it J. Phys. Soc. Japan }{\bf 72}, 735 (2003).

\bibitem{Ouyang04}
Z. Ouyang, C. Wang, and Z. S. She,
{\it Phys. Rev. Lett. }{\bf 93}, 078103 (2004). 

\bibitem{next} 
P. Messer, M. L\"assig, and P. Arndt, to be published. 

\bibitem{Eichler02a}
R. V. Samonte and E. E. Eichler,
{\it Nat. Rev. Genet. }{\bf 3}, 65 (2002).

\bibitem{Hsieh03}
L.-C. Hsieh, L. Luo, F. Ji, and H. C. Lee,
{\it Phys. Rev. Lett. }{\bf 90}, 018101 (2003).

\bibitem{Goffeau04}
A. Goffeau,
{\it Nature }(London) {\bf 430}, 25 (2004).

\bibitem{Lercher03}
M. J. Lercher, A. O. Urrutia, A. Pavlicek, and L. D. Hurst,
{\it Human Mol. Genetics }{\bf 12}, 2411 (2003).

\bibitem{Li91}
W. Li,
{\it Phys. Rev. A }{\bf 43}, 5240 (1991).

\bibitem{Mansilla00}
R. Mansilla and G. Cocho,
{\it Complex Systems }{\bf 12}, 207 (2000).

\bibitem{Durbin98}
R. Durbin, S. Eddy, A. Krogh, and G. Mitchison,
{\it Biological Sequence Analysis }
(Cambridge University Press, Cambridge, England, 1998).

\bibitem{Arndt04}
P. F. Arndt and T. Hwa,
{\it Bioinformatics }{\bf 20}, 1482 (2004).

\bibitem{Mouse02}
Mouse Genome Sequencing Consortium,
{\it Nature }(London) {\bf 420}, 520 (2002)

\bibitem{Eichler02b}
J. A. Bailey {\it et al.},
{\it Science }{\bf 297}, 1003 (2002)

\bibitem{Thomas04}
E. E. Thomas {\it et al.},
{\it PNAS }{\bf 101}, 10349 (2004)

\end{references}
\end{document}